\def\ts     {\thinspace}
\def\kms    {\ifmmode{{\rm \ts km\ts s}^{-1}}\else{\ts km\ts s$^{-1}$}\fi}
\def\mo     {M$_{\sun}$}

\def\hi     {H{\small I}}   
\def\hii    {H{\small II}}   
\def\ci     {C{\small I}}   
\def\cii    {C{\small II}}

\documentstyle[graphics]{mn}

\newif\ifAMStwofonts

\ifoldfss
  \ifCUPmtlplainloaded \else
    \NewTextAlphabet{textbfit} {cmbxti10} {}
    \NewTextAlphabet{textbfss} {cmssbx10} {}
    \NewMathAlphabet{mathbfit} {cmbxti10} {} 
    \NewMathAlphabet{mathbfss} {cmssbx10} {} 
  \fi
  \ifAMStwofonts
    \ifCUPmtlplainloaded \else
      \NewSymbolFont{upmath} {eurm10}
      \NewSymbolFont{AMSa} {msam10}
      \NewMathSymbol{\upi}     {0}{upmath}{19}
      \NewMathSymbol{\umu}     {0}{upmath}{16}
      \NewMathSymbol{\upartial}{0}{upmath}{40}
      \NewMathSymbol{\leqslant}{3}{AMSa}{36}
      \NewMathSymbol{\geqslant}{3}{AMSa}{3E}

       \let\le=\leqslant
       \let\ge=\geqslant
    \fi
  \fi
\fi 

\ifnfssone
  \newmathalphabet{\mathit}
  \addtoversion{normal}{\mathit}{cmr}{m}{it}
  \addtoversion{bold}{\mathit}{cmr}{bx}{it}
  \newmathalphabet{\mathbfit} 
  \addtoversion{normal}{\mathbfit}{cmr}{bx}{it}
  \addtoversion{bold}{\mathbfit}{cmr}{bx}{it}
  \newmathalphabet{\mathbfss} 
  \addtoversion{normal}{\mathbfss}{cmss}{bx}{n}
  \addtoversion{bold}{\mathbfss}{cmss}{bx}{n}
  \ifAMStwofonts
    \ifCUPmtlplainloaded \else
      \UseAMStwoboldmath
      \makeatletter
      \new@mathgroup\upmath@group
      \define@mathgroup\mv@normal\upmath@group{eur}{m}{n}
      \define@mathgroup\mv@bold\upmath@group{eur}{b}{n}
      \edef\UPM{\hexnumber\upmath@group}
      \new@mathgroup\amsa@group
      \define@mathgroup\mv@normal\amsa@group{msa}{m}{n}
      \define@mathgroup\mv@bold\amsa@group{msa}{m}{n}
      \edef\AMSa{\hexnumber\amsa@group}
      \makeatother
      \mathchardef\upi="0\UPM19
      \mathchardef\umu="0\UPM16
      \mathchardef\upartial="0\UPM40
      \mathchardef\leqslant="3\AMSa36
      \mathchardef\geqslant="3\AMSa3E

       \let\le=\leqslant
       \let\ge=\geqslant
    \fi
  \fi
\fi 

\ifnfsstwo
  \DeclareMathAlphabet{\mathbfit}{OT1}{cmr}{bx}{it}
  \SetMathAlphabet\mathbfit{bold}{OT1}{cmr}{bx}{it}
  \DeclareMathAlphabet{\mathbfss}{OT1}{cmss}{bx}{n}
  \SetMathAlphabet\mathbfss{bold}{OT1}{cmss}{bx}{n}
  \ifAMStwofonts
    \ifCUPmtlplainloaded \else
      \DeclareSymbolFont{UPM}{U}{eur}{m}{n}
      \SetSymbolFont{UPM}{bold}{U}{eur}{b}{n}
      \DeclareSymbolFont{AMSa}{U}{msa}{m}{n}
      \DeclareMathSymbol{\upi}{0}{UPM}{"19}
      \DeclareMathSymbol{\umu}{0}{UPM}{"16}
      \DeclareMathSymbol{\upartial}{0}{UPM}{"40}
      \DeclareMathSymbol{\leqslant}{3}{AMSa}{"36}
      \DeclareMathSymbol{\geqslant}{3}{AMSa}{"3E}

       \let\le=\leqslant
       \let\ge=\geqslant
    \fi
  \fi
\fi 

\ifCUPmtlplainloaded \else
  \ifAMStwofonts \else 
    \def\upi{\pi}
    \def\umu{\mu}
    \def\upartial{\partial}
  \fi
\fi

\title[Molecular gas in BCDGs]
{Molecular gas in blue compact dwarf galaxies}

\author[L.T. Barone et al.]
{L.T. Barone
   \thanks{\emph{Present address:} Osservatorio Astronomico di 
    Brera-Merate, Via Bianchi 46, I-23807 Merate (LC), Italy.}
   $^{1,2} $,
A. Heithausen$^1$, S. H\"uttemeister$^1$, T. Fritz$^1$
and U. Klein$^1$\\
$^1$Radioastronomisches Institut der Universit\"at Bonn, Auf
dem H\"ugel 71, D-53121 Bonn, Germany\\
$^2$Dipartimento di Astronomia, Universit\`a degli Studi di Bologna, Bologna, 
Italy.}

\date{Accepted .... Received ....}
\pagerange{\pageref{firstpage}--\pageref{lastpage}}
\pubyear{2000}

\def\LaTeX{L\kern-.36em\raise.3ex\hbox{a}\kern-.15em
    T\kern-.1667em\lower.7ex\hbox{E}\kern-.125emX}

\begin{document}

\label{firstpage}

\maketitle

\begin{abstract}

Blue compact dwarf galaxies (BCDGs) are currently undergoing strong bursts
of star formation. Nevertheless, only a few of them have been clearly detected
in CO, which is thought to trace the "fuel" of star formation: H$_2$. In this
paper we present a deep search for CO $J=1\rightarrow0$ and $J=2\rightarrow1$
emission lines in a sample of 8 BCDGs and two companions. Only 2 of them
(Haro\,2 and UM\,465) are detected. For the other galaxies we have obtained
more stringent upper limits on the CO luminosity than  published values. We
could not confirm the previously reported ``detection'' of CO for the
galaxies UM\,456 and UM\,462. We analyze a possible  relation between
metallicity, CO luminosity, and absolute blue magnitude of the galaxies. We
use previously determined
 relations between $X\equiv N$(H$_2)/I_{\rm CO}$ and the
metallicity to derive molecular cloud masses or upper limits for them. With
these ``global'' $X_{\rm CO}$-values we
find that for those galaxies which we detect in CO, the molecular gas mass is
similar to the \hi\ mass, whereas for the non-detections, the
upper limits on the molecular gas masses are significantly lower than the \hi\
mass. 
Using an LVG (Large Velocity Gradient) model we show that $X_{\rm CO}$ depends
not only on metallicity but also on other physical parameters such as, 
volume density and kinetic temperature, which rises the question on the
validity of ``global'' $X_{\rm CO}$-factors.

\end{abstract}

\begin{keywords}
 Galaxies: dwarf -- galaxies: ISM -- galaxies: starburst --
radio lines: ISM  
\end{keywords}


\begin{table*}
\caption{Properties of the galaxies and final observational results.}
\begin{tabular}{l c c c c c c c c c c }
\hline
 Galaxy & $\alpha $ & $\delta $ &  Distance & Abundance & $v_{hel}$ &
$\Delta v$ & rms (1 $\sigma$) & $I_{\rm CO}$ &
$L_{\rm CO}$ \\
 & (1950) & (1950) &  Mpc & 12+log\,O/H & {\scriptsize [\kms]}
& {\tiny [\kms]} & {\tiny [mK]} & {\tiny [K \kms]}
& {\tiny [10$^6$ K \kms pc$^2$]} \\
\hline
{UM\,422} & 11 17 37 &  +02 48 16 & 21.3 & $8.03$ 
& 1600 & -- & 3.6 & $<0.21$ & $<4.24 $\\
              &     &   &   &  & 1600 & -- & 7.5 & $<0.42$ & $<2.76 $ \\
\hline
{UM\,439} & 11 34 02 &  +01 05 38 & 14.7 & $8.03$ 
& 1097 & -- & 4.6 & $<0.27$ & $<1.83 $\\
         &       &   &      &  & 1097 & -- & 11.6 & $<0.64$ & $<1.56 $ \\
\hline
{UM\,446} & 11 39 02 & --01 37 26 & 24.0 & -- 
& 1792 & -- & 21.9 & $<0.43$ & $<2.22$ \\
         &          &   &     &  & 1792 & -- & 44.7 & $<2.47$ & $<4.01$ \\
\hline
{UM\,452} & 11 44 26 & --00 00 57 & 19.2 & -- 
& 1439 & -- & 7.7 & $<0.44$ & $<1.47$ \\
         &     &  &       &  & 1439 & -- & 11.6 & $<0.64$ & $<0.69$ \\
\hline
{UM\,456} & 11 48 01 & --00 17 23 & 23.3 & $7.89$ 
& 1749 & -- & 4.3 & $<0.25$ & $<3.85$ \\
         &          &  &     &  & 1749 & -- & 8.7 & $<0.47$ & $<2.78$ \\
\hline
 {UM\,456A} & 11 48 00 & --00 15 30 & 23.3 & -- & 1749
& -- & 14.5 & $<0.85$ & $<4.14$  \\
         &        &    &      &  & 1749 & -- & 27.1 & $<1.49$ & $<2.33$ \\
\hline
{UM\,456B} & 11 47 53 & --00 17 00  & 23.3 & -- 
& 1749 & -- & 9.1 & $<0.52$ & $<2.55$ \\
         &    &  &          &  & 1749 & -- & 18.2 & $<1.00$ & $<1.55$ \\
\hline
{UM\,462}  & 11 50 13 & --02 11 26 & 13.9 & $7.97$ & 
1051 & -- & 7.8 & $<0.46$ & $<0.79$ \\
         &      &  &    &   & 1051 & -- & 14.4 & $<0.80$ & $<0.44$ \\
\hline
{UM\,465} & 11 51 38 &  +00 24 56 & 15.4 & $8.57$ 
& 1144 & 17$\pm 2$ & 7.8 & 0.64$\pm0.07$  & 3.12$\pm0.32^b$ \\
     &      &  &    &   & 1144 & 35$\pm 13$& 
18.8 & 1.47$\pm0.35$ & 2.91$\pm0.71^b$ \\
\hline
{Haro\,2}$^a$ & 10 29 22 &  +54 39 24 & 20.5 & $8.4$
& 1401 & 19$\pm 8 $ & 9.3 & 4.2$\pm0.4$ & 36.7$\pm 3.5^b$ \\
              &          &            &      &     
& 1452 & 70.0 &      &     \\
              &          &            &      &  
& 1401 &18$\pm 20$ &  14.0 & 6.44$\pm1.33$ & 22.7$\pm4.7^b$\\
              &          &            &      &     
& 1452 & 70.0 &       &     \\
\noalign{\smallskip\hrule\smallskip
\noindent Remarks: First line for each galaxy refers to the ($J=1\to0$) 
transition, second line to the ($2\to1$) transition.
3 $\sigma$ upper limits to $I_{\rm CO}$ are obtained with
$\Delta v = 70$ \kms.
The distances are taken from Taylor et al.\ (1998) for
 most of the UM galaxies and from Loose \& Thuan (1986) for Haro\,2.
The distance of UM\,446 was obtained for a Hubble flow with
H$_0$=75 \kms Mpc$^{-1}$.
The velocities indicated in Column 7 are derived from H\,I
widths for the non-detections.
Metallicitites are taken from Campos-Aguilar, Moles \& Masegosa (1993)
for UM galaxies, who give a general uncertainty for the values of better than 
0.1 dex, and from Sage et al.\ (1992) for Haro\,2.
$a$: Two components found.
$b$: Errors include only statistical errors, not systematic ones.}
\end{tabular}
\label{osservazioni}
\end{table*}

\begin{table}
\caption{Observed positions}
\label{posizioni}
\begin{tabular}{l cccc}
\hline
Source & $\Delta\alpha^a$ & $\Delta\delta^a$ & Transition & rms \\
       &  [$\arcsec$] & [$\arcsec$]                  &            & [mK] \\
\hline
UM\,422 &   0 &  0  & $^{12}$CO(1-0) & 7 \\
        &     &     & $^{12}$CO(2-1) & 18 \\
        & --20& 20  & $^{12}$CO(1-0) & 7 \\
        &     &     & $^{12}$CO(2-1) & 13 \\
        & 20, &  0  & $^{12}$CO(1-0) & 8 \\
        &     &     & $^{12}$CO(2-1) & 22 \\
        & --20&  0  & $^{12}$CO(1-0) &  4 \\
        &     &     & $^{12}$CO(2-1) &  4 \\
        & 0   & 20  & $^{12}$CO(1-0) &  3 \\
        &     &      & $^{12}$CO(2-1) &  9 \\
\hline
UM\,439   & 0 &  0  & $^{12}$CO(1-0) & 5 \\
        &     &     & $^{12}$CO(2-1) & 18 \\
        & 10  &--20 & $^{12}$CO(1-0) & 5 \\
        &     &     & $^{12}$CO(2-1) & 13 \\
        & --10& 20  & $^{12}$CO(1-0) &  7 \\
        &     &     & $^{12}$CO(2-1) & 20 \\
        & 10  &  0  & $^{12}$CO(1-0) &  7 \\
        &     &     & $^{12}$CO(2-1) &  9 \\
\hline
UM\,446   &  0&  0  & $^{12}$CO(1-0) & 16 \\
        &     &     & $^{12}$CO(2-1) & 22 \\
\hline
UM\,452   & 0 &  5  & $^{12}$CO(1-0) &  5 \\
        &     &     & $^{12}$CO(2-1) &  7 \\
\hline
UM\,456   & 0 &  0  & $^{12}$CO(1-0) & 8 \\
        &     &     & $^{12}$CO(2-1) & 13 \\
        & 10  & 10  & $^{12}$CO(1-0) & 5 \\
        &     &     & $^{12}$CO(2-1) &  9 \\
        & 10  &  0  & $^{12}$CO(1-0) &  4 \\
        &     &     & $^{12}$CO(2-1) &  9 \\
        & 20  & 20  & $^{12}$CO(1-0) & 7 \\
        &     &     & $^{12}$CO(2-1) & 11 \\
UM\,456 A &  5&--5  & $^{12}$CO(1-0) & 11 \\
        &     &     & $^{12}$CO(2-1) & 13 \\
UM\,456 B &  0&  0  & $^{12}$CO(1-0) & 7 \\
        &     &     & $^{12}$CO(2-1) & 11 \\
\hline
UM\,462   & 10&  5  & $^{12}$CO(1-0) & 5  \\
        &     &     & $^{12}$CO(2-1) & 11 \\
\hline
UM\,465   &  2&  0  & $^{12}$CO(1-0) & 4 \\
        &     &     & $^{12}$CO(2-1) & 8 \\
        & --18& 10  & $^{12}$CO(1-0) & 4 \\
        &     &     & $^{12}$CO(2-1) & 9 \\
        &  --8& 10  & $^{12}$CO(1-0) & 7 \\
        &     &     & $^{12}$CO(2-1) & 11 \\
\hline
Haro\,2  &   0&  0  & $^{12}$CO(1-0) & 18 \\
        &     &     & $^{12}$CO(2-1) & 24 \\
        &  10 &  0  & $^{12}$CO(1-0) & 20 \\
        &     &     & $^{12}$CO(2-1) & 31 \\
        &   0& 10  & $^{12}$CO(1-0) & 19 \\
        &    &      & $^{12}$CO(2-1) & 33 \\
\hline
\noalign{\noindent Remarks: 
$a$: Offsets are relative to positions in Table \ref{osservazioni}.}
\end{tabular}
\end{table}

\section{Introduction \label{introduction}}

A particular class 
of dwarf galaxies named Blue Compact Dwarf Galaxies (BCDGs, Sargent \& Searle,
\shortcite{sargent:searle70})
has seen increasing interest among astrophysicists because of their 
extreme current star forming activity which is in contrast to their 
apparent ``youth'' in terms of chemical evolution.
BCDGs represent about 5\% of all dwarfs \cite{salzer89}, \cite{sage:etal92}
and are among the smallest star forming
galactic systems known.
\par
One of their outstanding properties is that their optical spectra are 
dominated by lines characteristic of \hii\ regions, which is the reason
why they are frequently termed ``\hii\ galaxies''. 
From optical spectroscopy we know that many \hii\ galaxies
have low heavy element abundances, typically down by a factor of three
up to more than twenty compared to the solar neighbourhood 
\cite{kunth:oestlin2000}.

It quickly became clear that these objects must form stars in what is 
called a burst, otherwise the observed star formation rate  would be in
conflict with their total gas masses as derived from \hi\ observations
\cite{thuan:martin81}. Such a burst may last some
$10^8$\,yrs, with a time span between bursts of the order of 10$^9$ yrs. It
has been suggested that interaction with companions might trigger their star
formation \cite{brinks90}, but Taylor et al. \shortcite{taylor:etal95}
found that only about 60\% of \hii\ galaxies have companions, often with
masses about $\frac{1}{10}$ of the main galaxy.

One of the most interesting and important issues which has not been settled so
far is the molecular gas content of these galaxies. Molecular hydrogen is
believed to be the preponderant seed for star formation, so it is a natural
assumption that large amounts of H$_2$ should be present in BCDGs. Yet the
results have been anything but conclusive so far. Following early attempts to
detect the CO line in BCDGs \cite{tacconi:young84}, there
have been a number of observations to confirm or reject those inconclusive
measurements (e.g. Sage et al. \shortcite{sage:etal92}, hereafter SSLH; Gondhalekar
et al. \shortcite{gondhalekar:etal98}; 
Taylor, Kobulnicky \& Skillman \shortcite{taylor:etal98}, hereafter
TKS). Surprisingly, the results remained partially contradictory, as for
instance in the case of II~Zw~40 (Arnault et al. \shortcite {arnault:etal88}; 
SSLH),  although improved instrumentation had been involved.

Prompted by the difficulty to detect the CO line -- relied upon as a good
tracer of molecular hydrogen content -- in BCDGs, some of the pertinent
publications prematurely concluded that molecular gas is deficient in these
systems. However, part of the difficulties to detect CO might have been 
due to beam filling and sensivitity problems. 
Taking e.g. 30\,Dor in the LMC as a
template giant star-forming complex, it is clear that, if placed 
at some larger distance
and covered by the beam, it could have escaped detection, as CO brightness
is rather low there, due to strong photo-dissociation in the high
interstellar radiation field 
\cite{cohen:etal88}. The same might be true for BCDGs. In this case, 
high-sensitivity mapping might reveal previously undetected CO emission.

We have therefore conducted a search for CO in gas-rich (based on \hi) \hii\
galaxies, using the IRAM 30\,m telescope. In contrast to previous projects
(e.g. SSLH, Gondhalekar et al. \shortcite{gondhalekar:etal98}), our observations
not only consisted of single pointings towards the brightest position in the
galaxies, but involved  mapping a number of positions in them, to detect
possible gas concentrations away from photodissociation regions.
Obtaining sensitive upper limits to the CO luminosity in these systems
is as much a goal of this study as detecting emission.

In Section 2 
we present details of our observations. In Section 3 
we present our data, and compare it with previous
results. In the subsequent section (Sec. 4) 
we discuss possible
causes for the detections and non-detections. This section is divided into
three subsections: Section \ref{gas} deals with the physical
conditions of the gas derived from a LVG model;
in Section \ref{metlum} we analyze the relationship between
metallicity and CO luminosity; finally, in Section
\ref{factor} we discuss the $X_{\rm CO}$ factor problem, which has
been heatedly debated in the past years and has not yet found a clear
resolution. Our conclusions are presented in the last section (Sec. 5).

\section{Observations \label{observations}}

The CO observations were carried out in October 1996 with the IRAM 30\,m 
Telescope on Pico Veleta in Spain. Our target galaxies -- except Haro\,2 --
have been selected from the sample examined by 
\cite{taylor:etal95} in the 21\,cm line of atomic hydrogen. 
Pointing positions were chosen from peak  \hi\ column densities, as seen in
their high angular resolution VLA observations.
The basic properties of the galaxies are listed in Table \ref{osservazioni}.
In the $J=1\rightarrow0$ transition (115 GHz) the HPBW is 22$''$, whilst in
the $J=2\rightarrow1$ transition (230 GHz) it is $12.\!''$5. 
At the distance of the galaxies, 14.7\,Mpc to 23.3\,Mpc, 
the beam size at 115\,GHz corresponds to 1.6\,kpc to 2.5\,kpc.

Two independent SIS
receivers have been used simultaneously at each frequency. More than 90\% of
our observations had $T_{sys}\le$ 600 K both at 230 and at 115 GHz. An
autocorrelator with a spectral resolution of 0.625 MHz at 115 MHz and 1.25 MHz
at 230 MHz and a filter spectrometer with 1 MHz resolution were used. A
baseline of zeroth or first 
order was always subtracted, and the spectra were summed up
to improve the signal-to-noise ratio; the spectra were finally smoothed to
roughly the same velocity resolution (5.2 \kms\ for the $J=1\rightarrow0$ and
4.8 \kms\ for the $J=2\rightarrow1$ transition).
Spectra were obtained with a wobbling secondary mirror with a wobbler 
throw of $\pm4'$ in azimuth.
All temperatures throughout this article refer to main beam brightness
with $T_{mb}$ derived using $\eta_{mb}(115)=0.74$ and
$\eta_{mb}(230)=0.45$.

\section{Results \label{results}}

Table \ref{osservazioni} gives an overview of the observed galaxies and some
of their known properties, and lists results of the galaxy-averaged CO spectra.
We mapped most of the galaxies in our sample in order to cover most of the
area where emission (perhaps in ``hot spots'') might be present, and obtained
low rms noise levels. Nevertheless, we detected CO only in Haro\,2 and in
UM\,465. For all the other galaxies we obtained low upper limits. 
It is interesting to note that the SMC, with an H$_2$ mass of 
$3\cdot10^7$\,\mo\ would be just detectable with our sensitivity if placed
at a distance of 15\,Mpc.

In Table \ref{posizioni}, all positions observed in all galaxies are listed,
with the final rms obtained for single positions. 
In Fig. \ref{haromap} and
\ref{um465map}, the $J=1\rightarrow0$ maps of Haro\,2 and UM\,465 are
displayed.

In Table \ref{osservazioni} we also give a summary of our results. For the
detections, we give the parameters of a Gaussian fit which we obtained for the
lines averaged over all positions. In all other cases, upper limits to the CO
intensity were calculated, again based on the average of {\it all} positions.
These upper limits were derived as $I_{\rm CO}\le \sigma \Delta
v_{\mbox {\tiny ch}} \sqrt{N}$, where $\sigma$ is the rms noise level obtained
in the baseline range, $\Delta v_{\mbox {\tiny ch}}$ is the velocity width of
each channel and $N$ is the number of channels involved. We always assumed a
total line width of 70 \kms, as this is the average velocity width one would
expect from the CO detections in the literature (see for example SSLH and
Gondhalekar et al. {\shortcite{gondhalekar:etal98}}). In the case of detections,
$I_{\rm CO}$ has been calculated based on Gaussian fits to the
spectra. Based on the $I_{\rm CO}$ values, CO luminosities ($L_{\rm CO}$) were
calculated. Those for UM\,465 and Haro\,2 are lower limits, because we did not
completely map the CO gas. The other values are upper limits for the galaxies.

\begin{table}
\caption{ Fit results for every position in Haro\,2 and UM\,465.} 
\label{fit}
\begin{tabular}{l  c c c c  c }
\hline
Galaxy & $\Delta\alpha$ & $\Delta\delta$ & 
$I_{\rm CO}$ & $v_{hel}$ & $\Delta v$   \\
 &  [$''$] & [$''$] & [K \kms] & [\kms] & [\kms]  \\
\hline
 Haro\,2 & 00 &00 & $3.8\pm0.3$ & $1452\pm2$ & $61\pm4$  \\
         &    &   & $6.2\pm0.4$ & $1455\pm2$ & $85\pm5$  \\
 & 10 & 00 & $3.9\pm0.4$ & $1441\pm4$ & $73\pm8$  \\
 &    &    & $3.7\pm0.7$ & $1443\pm6$ & $64\pm11$ \\
 & 00 & 10 & $4.2\pm0.2$ & $1445\pm2$ & $71\pm4$  \\
 &    &    & $7.6\pm0.3$ & $1446\pm2$ & $80\pm4$  \\
\hline
 UM\,465 & --08 & 10 & $0.64\pm0.13$  & $1143\pm2$ & $20\pm6$  \\
         &      &    & $0.95\pm0.13$ & $1147\pm1$ & $19\pm4$  \\
         & 02   & 00 & $0.31\pm0.07$ & $1144\pm1$ & $14\pm3$ \\
         &      &    & $0.60\pm0.07$ & $1140\pm6$ & $43\pm11$  \\
         & --18 & 10 & $0.75\pm0.15$ & $1153\pm6$ & $67\pm9$  \\
         &      &    & no line  \\
\hline
\noalign{Remarks: First line for each galaxy and position refers to the
($J=1\to0$) transition, second line to the ($2\to1$) transition.}
\end{tabular}
\end{table}

In the following sections we sumarize briefly our main results for the
 individual galaxies in comparison to previously published data where 
applicable. We also give a short description of the galaxies as they appear on
the Digitized Sky Survey (DSS).

\subsection{Haro\,2} 
Haro\, 2 is a relatively well studied BCDG. Its metallicity is about
$\frac{1}{3}$ solar. It has the shape  and the brightness profiles of an
elliptical galaxy, but possesses a brilliant blue nucleus which shows intense
star formation. The absolute blue magnitude is M$_B=-18.\!^{\rm m}4$ 
\cite{loose:thuan86}. A comparison between the UV, optical and FIR
spectra of Haro\,2 with evolutionary population synthesis models has allowed
to estimate the age of the youngest star formation episode to be 4 million
years, followed by two older bursts, the younger of which was over 20 million
years ago. 

\begin{figure}
\rotatebox{-90}{
\resizebox{8.cm}{!}{\includegraphics{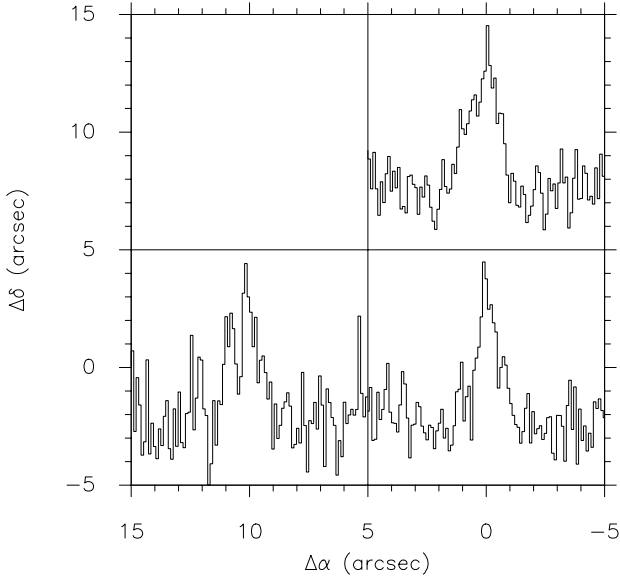}}}
\caption{Map of the $J=1\rightarrow0$ spectra of Haro\,2. 
The velocity range
is from 1200 to 1700 \kms. The temperature range is from $-0.03$ to 0.08 K.
Offsets are expressed in arcseconds relative to the position listed in
Tab. \ref{osservazioni}.}
\label{haromap}
\end{figure}

\begin{figure}
\resizebox{8cm}{!}{\includegraphics{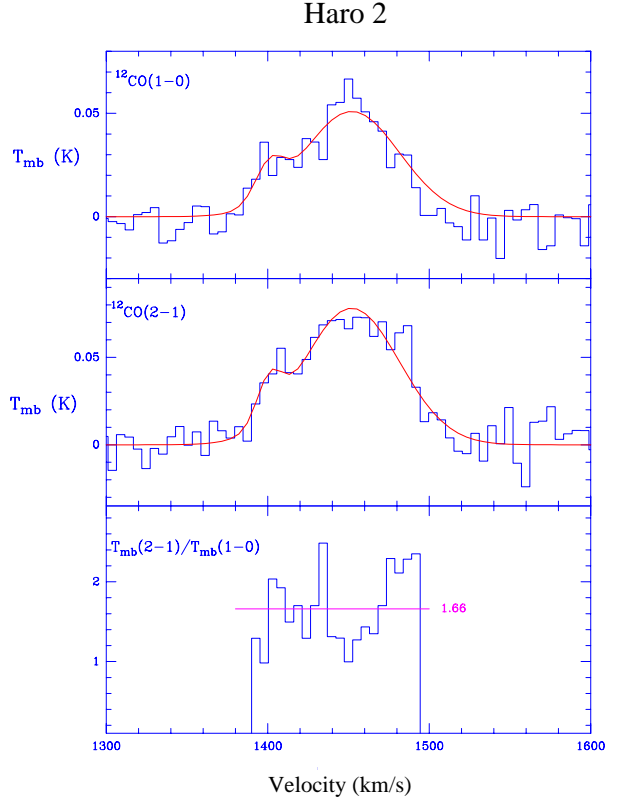}}
\caption{Average spectra and line ratio of Haro\,2. 
The presence of two components of the line is suggested, but it is not
reflected by a change of the line ratio.}
\label{haroratio}
\end{figure}

Our CO observations confirm the previous detection of CO in both the
$J=1\rightarrow0$ and $J=2\rightarrow1$ transitions by SSLH and by 
Knapp \& Rupen \shortcite{knapp:rupen96} 
in the latter transition. The emission is clearly
extended, as seen in Fig. \ref{haromap}, with significant lines  in all
positions (see Tab. \ref{fit}). Moreover, Fig. \ref{haroratio} suggests that
the line in Haro\,2 has two components which are seen at the same velocities
in both transitions. Recently, observations with the IRAM interferometer fully
confirmed this finding (Fritz et al., in prep.). The  line ratio of the
($2\to1$) to the ($1\to0$) line seems to be independent of the velocity (see
Fig. \ref{haroratio}). Following the path of SSLH, we calculate two extreme
line ratios, one assuming the source to be point-like, and one assuming a
uniformly filled beam. In the latter case, since a beam filling factor of 1 is
assumed, the ratio of the lines is effectively the ratio of integrated 
lines. This is also true if the filling factor is $< 1$, but
equal for both transitions (i.e. in the presence of  large scale clumping).
For a point source the other extreme is considered: one assumes that the
(2-1)/(1-0) intensity ratio is overestimated by exactly the ratio of the two
beam areas, so that the maximum line ratio must be divided by the ratio of the
squares of the two beam widths ($\simeq 3.1$). The results
are shown in Table \ref{ratios} and range from 0.5 to 1.5. The differences
between our ratios and those of SSLH might  be partly due to different main
beam  efficiencies used for the two transitions (not specified by SSLH).

\subsection{UM\,422}

The dominant \hii\ region of this dwarf galaxy is embedded in an
extended faint irregular stellar body. We have obtained 5 independent spectra;
none of them shows significant CO emission. UM\,422 has also been observed
with a single pointing by TKS using the NRAO 12\,m telescope. We confirm their
non-detection with a significantly lower upper limit.

\subsection{UM\,439}
In the optical, UM\,439 has a slightly elongated compact appearance with one
prominent \hii\ region south of the center. High resolution \hi\
observations by van Zee, Skillman \& Salzer \shortcite{vanzee:etal98} 
reveal that the star
formation is taking place in the peak of the extended gas distribution. It was
observed in CO by SSLH and by Gondhalekar et al. \shortcite{gondhalekar:etal98}.
We confirm their non-detections. Our upper limit to $I_{\rm CO}$ is a factor
of 2.4 lower than that of SSLH obtained with the NRAO 12\,m telescope and a
factor of 2.5 lower than that obtained with the Onsala 20\,m telescope.
Because we additionally observed 4 independent positions with higher angular
resolution than those just mentioned, our upper limit to $L_{\rm CO}$ is
significantly lower than those previously published.

\subsection{UM\,446}
The stellar component of this galaxy, as detected in optical imaging,
is very faint. We have 
observed it only in the central position. Our upper limit for CO 
is the first one ever published.

\subsection{UM\,452}
We have observed this galaxy towards one position only, where the optical 
emission is strongest. The optical extent of the galaxy looks much     
smaller as compared to the \hi. The mass of \hi\ is quite low,
$M_{HI}=5 \times 10^7$ \mo\  \cite{martin:99}.                                  
  
\subsection{UM\,456}
The star forming regions in this galaxy are confined to the center of
an extended and distorted stellar component.
Taylor et al. \shortcite{taylor:etal95} have detected two companions of UM\,456.
UM\,456~A seems to be a pure ``\hi\ cloud'' with no optical 
counterpart, whereas UM\,456~B is seen both in \hi\ and on optical
images. Both companions seem to be gravitationally bound to UM\,456.
None of them shows CO; with our better rms we do not confirm the
``marginal detection'' of UM\,456 by  SSLH; our
upper limit is a factor of 6 lower than their claimed detection.

\subsection{UM\,462}
The two BCDGs UM\,461 and UM\,462 seem to form a bound system with a linear
separation of about 70~kpc at a distance of 13.9~Mpc. Two centrally located
knots of star formation dominate the optical image of {UM\,462}. They are
associated with the peak of the \hi\  column density
\cite{vanzee:etal98}. The claimed detection of CO in UM\,462 by SSLH could
not be confirmed; our upper limit is  a factor of 2  lower. This galaxy had
also been observed by Gondhalekar et al. \shortcite{gondhalekar:etal98} with a
higher upper limit obtained at lower angular resolution.

\subsection{UM\,465}

\begin{figure}
\rotatebox{-90}{
\resizebox{6.5cm}{!}{\includegraphics{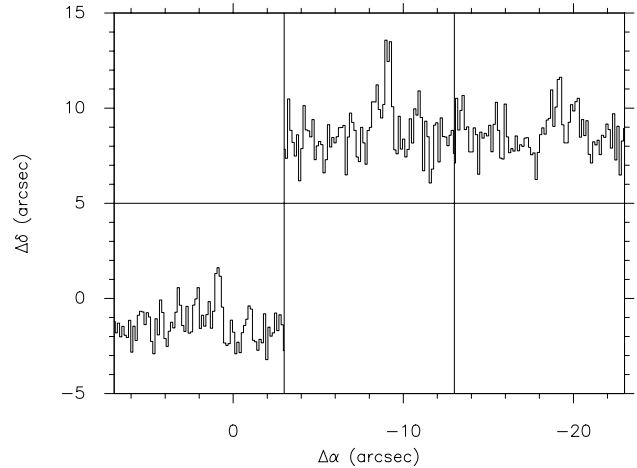}}}
\caption{Map of the $J=1\rightarrow0$ spectra of UM\,465. 
The velocity range
is from 900 to 1300 \kms. The temperatures range from $-0.02$ to 0.04 K.
Offsets are expressed in arcseconds relative to position given 
in Tab. \ref{osservazioni}}
\label{um465map}
\end{figure}

The optical appearance of this dwarf galaxy is circular in shape with an
exponential law brightness distribution 
\cite{doublier:etal97}. The nuclear starburst and extended dust lanes and
patches are well resolved by HST imaging \cite{malkan:etal98}.
A faint nearby object was not detected in \hi\  
\cite{taylor:etal95} but confirmed as a 
physical neighbour  \cite{doublier:etal97} using 6\,m telescope 
spectroscopy.  The HST images of
this companion reveal an irregular structure. While SSLH reported a marginal
detection in UM\,465 of CO, the present work delivers a clear one. The CO
emission is extended in this galaxy (see Fig. \ref{um465map}), but with a
lower intensity than Haro\,2. Only in one of the three positions no
$J=2\rightarrow1$ line was detected. The calculated line ratios are listed in
Table \ref{ratios}, and for this galaxy they range between 0.4 and 1.3.

\begin{table}
\caption{Comparison of the line intensity ratios of Haro\,2 and UM\,465,
calculated for different filling factors.}
\label{ratios}
\begin{tabular}{c c c c}
\hline
Source & Point source & Uniform filling & Reference \\
 & $\frac{2\rightarrow1}{1\rightarrow0}$ & 
$\frac{2\rightarrow1}{1\rightarrow0}$ & \\
 & & & \\
\hline
Haro\,2 & 0.31$\pm$0.04 & 1.0$\pm$0.1 & SSLH \\
       & 0.49$\pm$0.03 & 1.51$\pm$0.07 & this work$^a$ \\
UM\,465 & 0.27$\pm$0.11 & 0.91$\pm$0.36 & SSLH \\
       & 0.42$\pm$0.11 & 1.30$\pm$0.35 & this work \\
\noalign{\smallskip\hrule\smallskip
\noindent Remarks: The errors are derived formal errors of the fits. 
$a$: The two components have been averaged.}
\end{tabular}
\end{table}

\section{Discussion \label{discussion}}

Our deep observations corroborate the difficulty to detect CO in BCDGs. 
Only two of the galaxies observed show significant CO emission; in both cases it
is extended in both transitions. We note here that these
two sources are those with the highest metallicity in our sample. In the other
sources, even with observations towards several positions, we were unable to
find CO emission. One could be tempted to say that these galaxies are void of
molecular gas, but this conclusion is premature because the relationship  of CO
emission and H$_2$ content in a galaxy depends on many factors (Maloney \&
Black \shortcite{maloney:black88}, Israel \shortcite{israel97}), some of which
are not fully understood. Therefore, one can only conclude from the detection
of CO that H$_2$ is present, whereas the absence of CO does not necessarily
imply a lack of H$_2$.

\subsection{Physical conditions of the gas \label{gas}}

\begin{figure}
\rotatebox{-90}{
\resizebox{7.5cm}{!}{\includegraphics{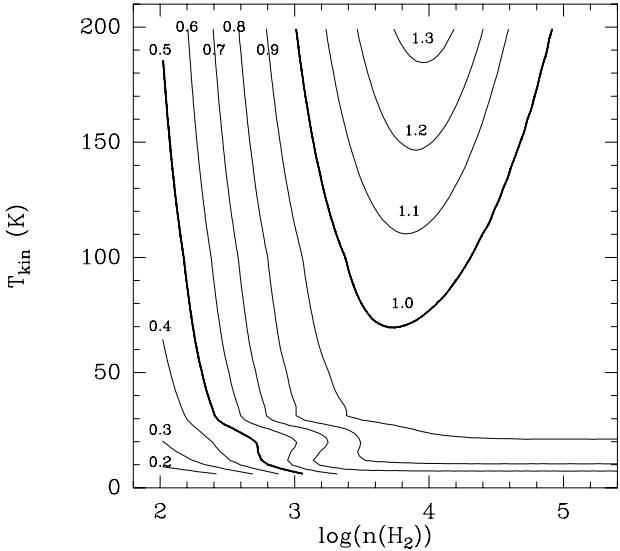}}}
\caption{Line intensity ratios as calculated with a LVG code. Solid 
lines represent $\cal R$$_{2/1}$, which is the predicted ratio between the 
$2\rightarrow1$ and the $1\rightarrow0$ lines.
A velocity gradient of
1 \kms pc$^{-1}$ and  [CO/H$_2]=2~10^{-5}$ have been assumed.
Contour values are given in the figure.}
\label{lvg}
\end{figure}

As a first step in our analysis, we try to infer some information on the
physical properties of the CO-emitting gas in BCDGs. We make use of a
large velocity gradient (LVG) model to predict ratios for the lowest CO lines.
The basic LVG assumption is that of a systematic monotonic
velocity gradient, which allows to treat the molecular excitation as a local
problem (see de Jong, Chu \& Dalgarno, 
 \shortcite{dejong:etal75} and White \shortcite{white77} 
for details). This is certainly an idealization of extra-galactic cloud 
complexes; however, an LVG code does not require detailed knowledge of 
the velocity field. As a first rough estimate this model further 
assumes constant density and kinetic
temperature within the molecular cloud. In Fig. \ref{lvg}, we show an example
of the dependence of line ratios on $n$(H$_2$) and $T_{kin}$. For this figure,
${{[{\rm CO/H}_2]}\over {|\nabla \vec{v}|}}=2\cdot 10^{-5}$ (\kms
pc$^{-1})^{-1}$ has been assumed. This corresponds to a velocity gradient of 1
\kms pc$^{-1}$ and an abundance of [CO/H$_2$]$=2~10^{-5}$ or, due to how the
LVG code is constructed, to a velocity gradient of 5 \kms pc$^{-1}$ and
[CO/H$_2$]=10$^{-4}$. The figure shows the ratios of the intensities of the
$2\rightarrow1$ to $1\rightarrow0$ transitions, $\cal R$$_{2/1}$, and that of 
$3\rightarrow2$ and the $2\rightarrow1$ lines, $\cal R$$_{3/2}$.

As discussed above, for Haro\,2 and UM\,465 the line ratios are  $0.49\le{\cal
R}_{2/1}\le1.51$ and  $0.42\le{\cal R}_{2/1}\le1.30$, respectively, depending
on the filling of the sources in our beam. From our limited mapping of the two
galaxies, we know that the sources are extended and the ``point source'' limit
can be firmly excluded. On the other hand, observations of Haro\,2 obtained
with the Plateu de Bure Interferometer (Fritz et al., in prep.) show that the
galaxy does not fill the beam of the IRAM 30\,m telescope uniformly. It is
then reasonable to expect a value for the line intensity ratios in between
those listed in Table \ref{ratios}, and may thus be close to unity.

Adopting a ratio ${\cal R}_{2/1}$ between 0.8 and 1.0, we can derive from Fig.
\ref{lvg} that the gas is either at high temperatures with medium densities
($\approx$ a few hundred cm$^{-3}$) or at high densities 
($\ge 2000$\,cm$^{-3}$)
 and low temperatures ($T_{kin}\le50$\,K). More stringent limits on volume
density and kinetic temperatures require observations of higher $^{12}$CO
transitions and/or $^{13}$CO transitions. These transitions are expected to
be very weak though, and  thus difficult and time-consuming.

\subsection{Dependence of the CO luminosity on
metallicity and absolute blue magnitude \label{metlum}}

\begin{figure*}
\rotatebox{-90}{
\resizebox{9cm}{!}{\includegraphics{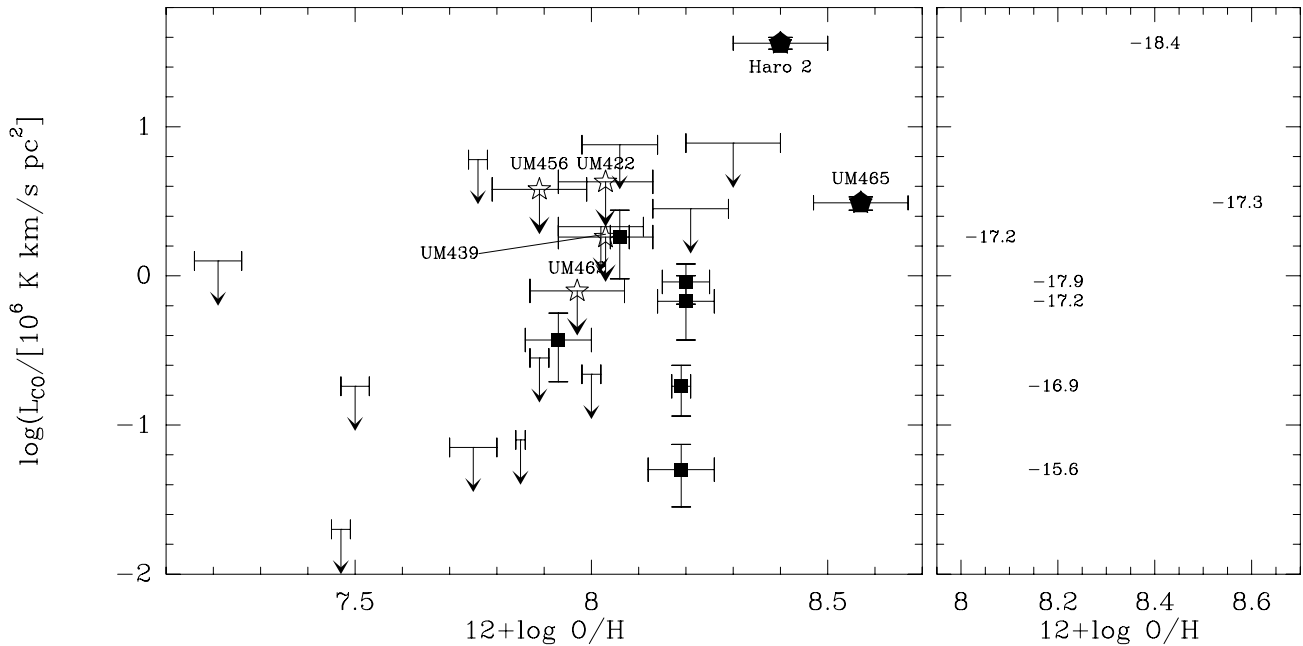}}}
\caption{Dependence of the CO luminosity on the metallicity of dwarf
galaxies. Our non-detections are marked as open stars, detections as
filled diamonds. Detections from the compilation of TKS 
are marked as filled squares. Arrows mark upper limits. On the left side
of the plot all galaxies are shown. On the right side only 
galaxies with a well determined CO luminosity are shown. 
Here the numbers represent the absolute blue magnitudes of the galaxies.}
\label{lumino}
\end{figure*}

We subsequently examine the relation of the CO emission and
metallicity. Because only
the two galaxies in our sample with the highest metallicities are detected in
CO, one could expect that CO luminosity depends on metallicity. Therefore,
we plot our CO luminosities ($L_{\rm CO}$) -- listed in Table
\ref{osservazioni} -- vs. the metallicities of the galaxies of our sample.
Since not all of the galaxies have known metallicities, we are left with
only 6 galaxies. These are shown in Fig. \ref{lumino}. In this figure we also
include the data points given by TKS because
they represent the most comprehensive sample of CO observations of
dwarf galaxies with metallicity
determinations. We only selected those dwarf galaxies from the sample with 
 metallicities better determined than 0.1 dex; 
these galaxies are, however, not necessarily classical BCDGs.

We have chosen the CO luminosity because it is largely independent of
distance, although the error in the distance determination enters as the 
square in
the CO luminosity. Furthermore, the CO luminosity is directly proportional to
the H$_2$ mass corrected for helium
($M({\rm H}_2)=2.2\cdot X_{\rm CO} \cdot L_{\rm CO}$), 
with $M$ in \mo\ and
$L_{\rm CO}$ in K\,\kms\,pc$^2$. $X_{\rm CO}$ (in units of $10^{20}$\,molecules
\,cm$^{-2}$\,(K\,\kms)$^{-1}$) is the well-known but poorly
determined $X_{\rm CO}$ factor which relates the molecular hydrogen column 
density to
the observed integrated CO line intensity ($X_{\rm CO}\equiv N({\rm
H}_2)/I_{\rm CO}$).

Fig. \ref{lumino} shows a general trend that galaxies with higher metallicities
have higher CO luminosities, although no functional correlation is visible.
TKS have recently proposed that galaxies with metallicities below
7.9 are basically undetectable in CO. Our data do not contradict this finding.
For galaxies close to that limit we were only able to derive upper limits.
We note, however, that even above that limit there are galaxies not 
detected in CO.

Although it is qualitatively expected that a higher metallicity leads to
a higher $L_{\rm CO}$ because of the availability of the building 
blocks of the CO molecule, Fig. \ref{lumino} shows that the oxygen
abundance cannot be the only factor influencing the CO luminosity. Because
BCDGs are actively star-forming galaxies, the UV radiation
field may be locally high.
This plays two conflicting roles for CO: on the one hand, it heats the gas,
so that the excitation temperature is higher and CO is brighter; on the other
hand, if hard enough,  it destroys CO via photodissociation, thus the CO
emission becomes weaker. Pak et al. \shortcite{pak:etal98} and 
Bolatto, Jackson \& Ingalls 
\shortcite{bolatto:etal99} have studied these effects and found that the CO
emitting regions are effectively smaller in low-metallicity environments and
most of the carbon is present in atomic form. The net
effect in galaxies with low metallicities is that, due to low beam filling of
the clouds, the molecular gas becomes invisible in the CO lines and  might be
better traced in the fine structure lines of \ci\ and \cii. This different beam
filling might be the cause for the dependence of the integrated CO line
intensity $I_{\rm CO}$\ on the metallicity found by TKS.

To study the effect of the radiation field on the CO luminosity, we labelled
the values in the CO luminosity-metallicity plane with the absolute blue
magnitudes of the galaxies (right side of Fig. \ref{lumino}). Again there is
no simple relation between any two of these quantities. However, there is
evidence for an influence of both metallicity and absolute blue magnitude on
the CO luminosity, meaning that higher blue magnitudes lead to higher CO
luminosity for a given metallicity. It also appears that at lower 
metallicities a higher absolute blue magnitude is necessary to reach a certain
CO luminosity. Clearly, more well-observed galaxies are necessary to study the
relation between the three quantities.

\subsection {Molecular gas masses and the $X_{\rm CO}$ factor\label{factor}}

Directly linked to the question of the dependence of the CO luminosity on
metallicity and radiation field is the question of which $X_{\rm CO}$ factor is
applicable to low-metallicity galaxies. A number of studies have examined
possible correlations between $X_{\rm CO}$ and the metallicity (e.g. Dettmar \&
Heithausen \shortcite{dettmar:heithausen89}, Wilson \shortcite{wilson95},
Verter \& Hodge \shortcite{verter:hodge95}, Arimoto, Sofue \& Tsujimoto 
\shortcite{arimoto:etal96}). Klein \shortcite{klein99} has proposed an
additional dependence of $X_{\rm CO}$ on the cosmic ray flux as judged  from
the radio continuum brightness.

An independent determination of the $X_{\rm CO}$ factor for our BCDG sample
would be  useful to determine their molecular masses; this is, however, beyond
the scope of this paper. A reliable value for $X_{\rm CO}$ has been
established for the disk of the Milky Way, $X_{\rm MW}$. The currently best
accepted value is $X_{\rm MW} = 1.6\cdot10^{20}$ molecules cm$^{-2}$
K$^{-1}$ \kms\ \cite{hunter:etal97}. $X_{\rm CO}$
factors determined for galaxies with lower metallicities are usually
significantly higher (e.g. Cohen et al. \shortcite{cohen:etal88}, Dettmar \&
Heithausen \shortcite{dettmar:heithausen89}).

\begin{table}
\caption {Molecular masses from predicted $X_{\rm CO}$ factors.} 
\label{xfactors}
\begin{tabular}{l l l l }
\noalign{\hrule\smallskip}
Source & $X_{\rm CO}$ & $M($H$_2$) & $M$(\hi) \\
       &       & $10^8$\mo & $10^8$\mo \\
\noalign{\smallskip}
\hline
UM\,422 & 18.6 & $\le1.3$ & 26\\
UM\,439 & 18.6 & $\le0.6$ & 3.5\\
UM\,456 & 25.7 & $\le1.6$ & 3.7\\
UM\,462 & 21.4 & $\le0.3$ & 2.9\\
UM\,465 &  5.4 & 0.3 & 0.39\\
Haro\,2 &  7.9 & 4.7 & 4.8\\ 
\noalign{\smallskip\hrule\smallskip
\noindent Remarks: 
$X_{\rm CO}$ factors are derived from the metallicity dependence 
as given by Arimoto et al. \shortcite{arimoto:etal96}, using metallicities
from Table \ref{osservazioni}. $X_{\rm CO}$ is given in units of
$10^{20}$ molecules cm$^{-2}$ (K\,\kms)$^{-1}$. \hi\ masses for UM galaxies 
are from Taylor et al. \shortcite{taylor:etal95}, for Haro\,2 from SSLH.}
\end{tabular}
\end{table}
       
In the following, we assume that the correlation between $X_{\rm CO}$, derived
from a virialization analysis of several galaxies,  and the 
oxygen abundance 
obtained by Arimoto et al. \shortcite{arimoto:etal96} gives $X_{\rm CO}$ 
factors applicable to our galaxy sample. 
H$_2$ masses
derived under this assumption are listed in Table \ref{xfactors}.
Also given are the \hi\ masses as derived by 
\cite{taylor:etal95} for the UM galaxies and by SSLH for Haro\,2.
One remarkable result from this calculation is that those
galaxies undetected in CO have upper limits on the molecular gas mass 
significantly below the \hi\ mass, whereas in the two galaxies where
CO is detected, \hi\ and H$_2$ masses are about the same.

The $X_{\rm CO}$ factors used here to derive molecular gas masses are 
{\it global}
factors. Studying the CO emission of the Magellanic Clouds with different
angular resolutions Rubio, Lequeux \& Boulanger \shortcite{rubio:etal93}
noted that the derived $X_{\rm CO}$ factor depends on
the linear resolution, implying that the {\it local} $X_{\rm CO}$
value is lower than the {\it global} one.

In order to calculate the {\it local} $X_{\rm CO}$ we use an LVG approximation 
assuming that H$_2$ and CO share the same volume. 
The LVG approximation can then be used to derive the H$_2$ column density,
$N({\rm H}_2)$, 
from  $N({\rm H}_2)={n({\rm H}_2) \Delta v}/{|\nabla \vec{v}|}$, 
where $n({\rm H}_2)$ is the H$_2$ volume density, $\Delta v$ is
the line width and $|\nabla \vec{v}|$ is the velocity gradient.
Since $I_{\rm CO}\simeq T_{mb} \Delta v$, one can derive 
$X_{\rm CO}$ from $X_{\rm CO}=N({\rm H}_2)/I_{\rm CO}\propto
{n({\rm H}_2)}/{|\nabla \vec{v}|T_{mb}}$.

\begin{figure}
\resizebox{8cm}{!}{\includegraphics{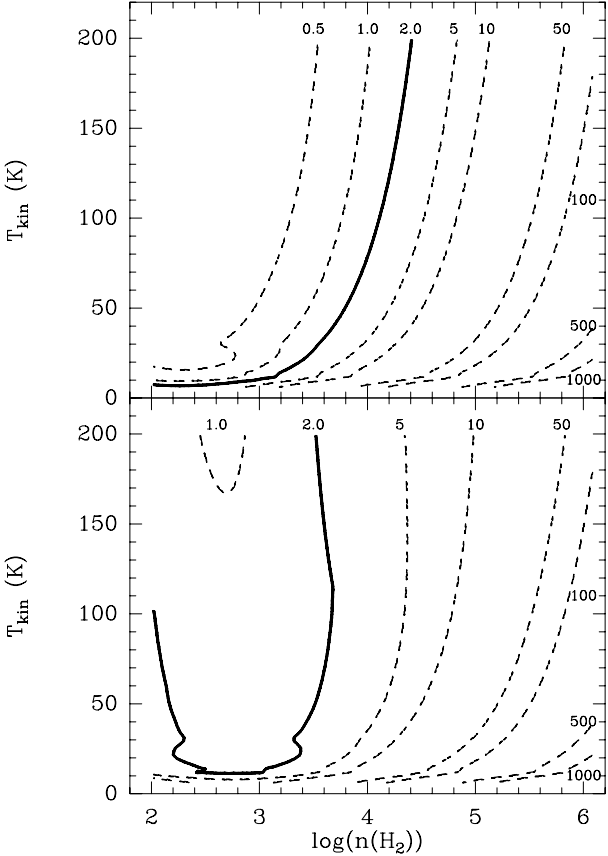}}
\caption{The variation of $X_{\rm CO}$ 
with changing kinetic temperature and volume density,
as derived from the LVG approximation. Values at the contours give 
$X_{\rm CO}$ in
units of $10^{20}$\,cm$^{-2}$\,(K\,\kms)$^{-1}$.
The velocity gradient is fixed to 2 \kms pc$^{-1}$. In the top panel,
the abundance is [CO/H$_2$]=10$^{-4}$; in the bottom panel [CO/H$_2$]=
10$^{-5}$, two values between which it is likely to find most BCDGs.
The thick solid lines represent  the value for the 
Milky Way, $X_{\rm MW}\sim 2\cdot10^{20}$\,cm$^{-2}$\,(K\,\kms)$^{-1}$.}
\label{xfile}
\end{figure}

The dependence of the local $X_{\rm CO}$ on varying volume density and
kinetic temperature is shown in Fig. \ref{xfile}.
The velocity gradient is fixed to 2 \kms pc$^{-1}$. In the top panel,
the abundance is [CO/H$_2$]=10$^{-4}$; in the bottom panel [CO/H$_2$]=
10$^{-5}$. These are extreme values for the metallicities of BCDGs.
$X_{\rm MW}$ is indicated by the thick solid line.
We note that, once the density becomes high enough, $X_{\rm CO}$ 
will not change significantly any more, due to the growth of the optical depth,
and $X_{\rm CO}$ is independent of the abundance. On the other
hand, at lower densities and kinetic temperatures, 
$X_{\rm CO}$ changes significantly depending on
$n$(H$_2$) and $T_{kin}$ for a given metallicity. 

{\it These simple calculations indicate that $X_{\rm CO}$  does not 
only depend on the metallicity of a galaxy.} 
Physical parameters, such as average volume density
and kinetic temperature, play important roles, if the density is below
$10^4$ cm$^{-3}$ and/or the kinetic temperature is below 50\,K, values
found in many molecular clouds in the Milky Way. 
The $X_{\rm CO}$-factor, that we find for a standard density of
$10^3$\,cm$^{-3}$ and a low-metallicity environment, is low, i.e. close to
Galactic. In contrast, $X_{\rm CO}$ derived from the formula of Arimoto et al.
\shortcite{arimoto:etal96} is higher by an order of magnitude. This supports
the concept of large amounts of hidden gas - either atomic or molecular.

Interferometric observations of BCDGs might help
to resolve this issue.  Such a study of the nearby
post-starburst dwarf NGC\,1569, a galaxy that may be considered as a
nearby BCDG in a post-starburst phase, yielded a rather high
value of $X_{\rm CO}=6.6 \cdot X_{\rm MW}$,
based on virial masses of resolved GMCs \cite{taylor:etal99}.
This indicates that  this method, which is also the basis of Arimoto's
formula, is sensitive to the `hidden' H$_2$ and tends to yield global
values for  $X_{\rm CO}$. In contrast, we expect to find local
values if  line ratio studies encompassing several CO isotopomers
become available, since these studies directly probe the physics
of the  gas from which the CO emission arises.

\section{Conclusions \label{conclusion}}

We have searched for emission from the 
$^{12}$CO ($J=1\rightarrow0$ and $J=2\rightarrow1$) 
transitions in 10 dwarf galaxies, 8 of which are BCDGs and 2 are the
companions of one of these. We detected CO in 2 of them (Haro\,2 and UM\,465)
and found it to be extended in both galaxies. Although we mapped part of the
other galaxies, we were unable to detected CO. We obtained very stringent
upper limits. We  could not confirm the ``marginal detection'' of CO in
UM\,456 and UM\,462 previously reported by SSLH.
 
The observed line ratios of the $2\rightarrow1$ to $1\rightarrow0$ transitions
are not very sensitive to changes in the kinetic temperature. Modelling
the ratio with a simple LVG code helps only to exclude low densities.
Higher CO transitions and/or observations of CO isotopomers will help to
get more stringent limits on these physical parameters.

We could not find any simple relation between metallicity and CO luminosity.
Molecular gas masses for the galaxies are derived assuming the relation 
between $X_{\rm CO}$ and metallicity given by \cite{arimoto:etal96}.
We find that for those galaxies detected in the CO lines the molecular gas mass
is comparable to the \hi\ mass, whereas for those galaxies undetected in CO
the \hi\ mass is significantly larger than the limits on the molecular 
gas mass.

Even in the sources where CO has not been detected, we do not argue against
the presence of H$_2$. While it is certainly possible that in the extreme 
environment of a BCDG not just CO but also H$_2$ is destroyed, at least 
in regions close to young massive stellar clusters, a picture in which
a large amount of H$_2$ exists without CO is attractive.
Sensitive observations of \ci\ and \cii\ in these galaxies would thus be
desirable in the future to shed light on this issue.

\section*{Acknowledgements}
{L.T.B. would like to thank Prof. Loretta Gregorini and the Socrates/Erasmus
project which made this exchange possible and financed it, the Faculty of
Science of the University of Bologna and the C.N.A.A. ({\it Consorzio Nazionale
per l'Astronomia e l'Astrofisica}) for grants supporting this work.
This project was supported by the Deutsche Forschungsgemeinschaft via
the Graduiertenkolleg {\it ``The Magellanic Clouds and other Dwarf 
Galaxies''.}}

\bsp

\label{lastpage}

\end{document}